\documentclass[prb,amsmath,amsfonts,amssymb]{revtex4}
\usepackage{graphicx}
\usepackage{bm}
\usepackage{color}

\newcommand{\braket}[2]{\left\langle \, #1 \, | \, #2 \right \rangle}
\newcommand{\brakett}[2]{\left\langle \, #1 \, \| \, #2 \right \rangle}
\newcommand{\A}{\mathbf{A}}
\newcommand{\B}{\mathbf{B}}

\newcommand{\R}{\mathbf{R}}
\renewcommand{\S}{\mathbf{S}}
\newcommand{\T}{\mathbf{T}}
\newcommand{\X}{\mathbf{X}}
\newcommand{\Y}{\mathbf{Y}}
\newcommand{\w}{\bm{\omega}}
\newcommand{\z}{\mathbf{z}}

\begin{document}
\title{The Ground State Correlation Energy of the Random Phase Approximation from a Ring Coupled Cluster Doubles Approach}
\author{Gustavo E. Scuseria}
\author{Thomas M. Henderson}
\affiliation{Department of Chemistry, Rice University, Houston, TX 77005-1892}
\author{Danny C. Sorensen}
\affiliation{Department of Computational and Applied Mathematics, Rice University, Houston, TX 77005-1892}
\date{\today}

\begin{abstract}
We present an analytic proof demonstrating the equivalence between the Random 
Phase Approximation (RPA) to the ground state correlation energy and a 
ring-diagram simplification of the Coupled Cluster Doubles (CCD) equations.  
In the CCD framework, the RPA equations can be solved in $\mathcal{O}(N^4)$ 
computational effort, where $N$ is proportional to the number of basis 
functions.
\end{abstract}
\maketitle

There has recently been a revival of interest in RPA in the quantum chemistry 
community.  The RPA is popular for calculations of excitation energies both in 
finite systems\cite{Oddershede,FurcheVanVoorhis} and in 
solids,\cite{Kresse,FuchsGonze} and is related to time-dependent density 
functional theory.\cite{Casida,Casida2,RubioRMP}  As a technique for 
describing electronic correlations, RPA has significant advantages, 
particularly for those interested in density functional theory.  It describes 
dispersion and van der Waals interactions correctly,\cite{Angyan,Dobson} and 
is exact for long-range correlations.\cite{YanPerdewKurth}  Left-right static 
correlations seem to be properly described by RPA,\cite{GonzeFuchsBurke} and 
RPA fixes the pathologies of nonlocal Hartree-Fock-type exchange in metallic 
systems.  Readers interested in details about RPA for ground state correlation 
can refer to the recent paper by Furche\cite{Furche} where he discusses an 
interesting simplification to reduce the computational cost of RPA correlation 
and provides ample background information about RPA.  Note that his work 
focuses on direct RPA, in which the exchange terms are neglected; as discussed 
later in this communication, this is the form of RPA most useful in the 
context of density functional theory.

A connection between the RPA correlation energy and a ring diagram 
approximation to CCD was first mentioned by Freeman in his 1977 
paper.\cite{Freeman}  Very recently, A. Gr\"uneis and G. Kresse reproduced this 
evidence and found numerical proof of the equivalence between these two 
approaches.\cite{Kresse2}  Here, we offer an analytic proof that these two 
problems yield identical correlation energies.  To the best of our knowledge, 
no such formal proof has been given before.

As a method for calculating electronic excitation spectra, RPA requires the 
solution of
\begin{equation}
\begin{pmatrix}
\hfill \A & \hfill \B \\  -\B & -\A
\end{pmatrix}
\begin{pmatrix}
\X \\ \Y
\end{pmatrix}
=
\begin{pmatrix}
\X \\ \Y
\end{pmatrix}
\w.
\label{RPA}
\end{equation}
The matrices $\A$, $\B$, $\X$, and $\Y$ are all $o v \times o v$, where $o$ 
and $v$ are respectively the number of occupied and unoccupied spin-orbitals.  
The eigenvalue problem above can be completed by noting that if 
$\bigl(\begin{smallmatrix} \X_i \\ \Y_i \end{smallmatrix}\bigr)$ is an 
eigenvector with eigenvalue $\omega_i$, then 
$ \bigl(\begin{smallmatrix} \Y_i \\ \X_i \end{smallmatrix}\bigr)$ is also an 
eigenvector, with eigenvalue $-\omega_i$.  In the (real) canonical spin-orbital 
basis we use throughout this letter, we have
\begin{subequations}
\begin{alignat}{1}
A_{ia,jb} &= (\epsilon_a - \epsilon_i) \delta_{ij} \delta_{ab} + \brakett{ib}{aj},
\\
B_{ia,jb} &= \brakett{ij}{ab}.
\end{alignat}
\label{DefAB}
\end{subequations}
Here, $\epsilon_p$ is a diagonal element of the Fock operator.  Indices $i$, 
$j$, $k$, and $l$ indicate occupied spin-orbitals, while $a$, $b$, $c$, $d$ 
indicate unoccupied spin-orbitals.  For arbitrary spin-orbitals $p$, $q$, 
$r$, and $s$, the two-electron integral $\brakett{pq}{rs}$ is defined by
\begin{subequations}
\begin{alignat}{1}
\brakett{pq}{rs} &= \braket{pq}{rs} - \braket{pq}{sr},
\\
\braket{pq}{rs}  &= \int \int \mathrm{d}\mathbf{x}_1 \, \mathrm{d}\mathbf{x}_2 \, \phi_p(\mathbf{x}_1) \, \phi_q(\mathbf{x}_2) \, \frac{1}{r_{12}} \, \phi_r(\mathbf{x}_1) \, \phi_s(\mathbf{x}_2)
\end{alignat}
\end{subequations}
where $\mathbf{x}$ is a combined space and spin electron coordinate.

The RPA correlation energy can be obtained by considering two harmonic 
excitation energy problems:\cite{Furche,RingSchuck} RPA and the Tamm-Dancoff 
approximation (TDA) thereto, which sets $\B = \bm{0}$ and thus solves
\begin{equation}
\A \, \mathbf{Z} = \mathbf{Z} \, \bm{\nu}.
\end{equation}
In the quantum chemistry community, TDA is also known as configuration 
interaction singles (CIS).  While TDA includes only excitation operators, RPA 
also includes de-excitation operators which can be thought of as correlating 
the ground state.  The ground state correlation energy in RPA is given by the 
difference between the zero point energies of these two harmonic oscillator
excitation problems with correlated (RPA) and uncorrelated (TDA) ground 
states.  We thus have
\begin{equation}
E_c^{RPA} = \frac{1}{2} \sum_i {}^\prime \left(\omega_i - \nu_i\right) = \frac{1}{2} \mathrm{Tr}(\w - \A).
\label{EcDRPA}
\end{equation}
The prime on the summation means that we include only the positive excitation 
energies in defining $\w$.

A different approach to calculating the correlation energy is coupled cluster 
theory.  The simplest coupled cluster method includes only double excitations 
from the ground state, and is termed CCD.  The CCD correlation energy is
\begin{equation}
E_c^{CCD} = \frac{1}{4} \sum \brakett{ij}{ab} t_{ij}^{ab} = \frac{1}{2} \sum \braket{ij}{ab} t_{ij}^{ab},
\label{EcCCD0}
\end{equation}
where in the last equation we have used antisymmetry of $t_{ij}^{ab}$ under 
interchange of $i$ with $j$ or $a$ with $b$.  To determine the $t_{ij}^{ab}$, 
we solve the CCD equations in the spin-orbital basis (see, for example, Ref. 
\onlinecite{GusCCD}),
\begin{equation}
\begin{split}
0 = &\brakett{ij}{ab} + (\epsilon_a + \epsilon_b - \epsilon_i - \epsilon_j) t_{ij}^{ab}  + \frac{1}{2} t_{kl}^{ab} \brakett{ij}{kl} + \frac{1}{2} t_{ij}^{cd} \brakett{ab}{cd} + \frac{1}{4} t_{kl}^{ab} \brakett{kl}{cd} t_{ij}^{cd} 
\\
    &- \frac{1}{2} P_{ab} t_{ij}^{cb} \brakett{kl}{cd} t_{kl}^{ad} - \frac{1}{2} P_{ij} t_{kj}^{ab} \brakett{kl}{cd} t_{il}^{cd} + P_{ij} P_{ab} t_{jk}^{bc} \left(\brakett{ic}{ak} + \frac{1}{2} \brakett{kl}{cd} t_{il}^{ad}\right),
\end{split}
\end{equation}
where internal indices ($k$, $l$, $c$, and $d$) are to be summed, and $P_{ij}$ 
and $P_{ab}$ are permutation operators: ($P_{ab} g_{ac} = g_{ac} - g_{bc}$, 
\textit{etc}.).  Keeping only particle-hole ring contractions, leads to what 
we shall here term ``ring-CCD'' (rCCD),
\begin{equation}
0 = \brakett{ij}{ab} 
+ t_{ik}^{ac} \left(\epsilon_c - \epsilon_k\right) \delta_{bc} \delta_{jk} 
+  \left(\epsilon_c - \epsilon_k\right) \delta_{ac} \delta_{ik} t_{kj}^{cb}
+ \brakett{ic}{ak} t_{kj}^{cb} + t_{ik}^{ac} \brakett{jc}{bk} + t_{ik}^{ac} \brakett{kl}{cd} t_{lj}^{db}.
\label{rCCD}
\end{equation}
Defining $t_{ij}^{ab} = T_{ia,jb}$, and using Eqn. \ref{DefAB}, we obtain
\begin{equation}
\B + \A \, \T + \T \, \A + \T \, \B \, \T = \bm{0}.
\label{Riccati}
\end{equation}
Removing the exchange integrals (\textit{i.e.} setting 
$\brakett{pq}{rs} \longrightarrow \braket{pq}{rs}$) in Eqn. \ref{rCCD} gives 
us what we will call direct ring-CCD (drCCD), and in Eqn. \ref{DefAB} gives us 
direct RPA.  Thus, Eqn. \ref{Riccati} holds both for rCCD and for direct rCCD 
with the $\A$ and $\B$ matrices defined as in RPA or direct RPA, respectively. 
In terms of $\B$ and $\T$, the rCCD correlation energy is
\begin{equation}
E_c^{rCCD} = \frac{1}{4} \mathrm{Tr}(\B \, \T),
\label{Ec_rCCD}
\end{equation}
while the drCCD correlation energy picks up an extra factor of two due to the 
different definition of $\B$:
\begin{equation}
E_c^{drCCD} = \frac{1}{2} \mathrm{Tr}(\B \, \T).
\label{Ec_drCCD}
\end{equation}

We prove here that Eqn. \ref{Riccati} can be obtained from the RPA equations, 
and that with $\T$ thereby defined, the direct rCCD correlation energy of Eqn. 
\ref{Ec_drCCD} is equal to the direct RPA correlation energy of Eqn. 
\ref{EcDRPA}.

We begin with the RPA equations, Eqn. \ref{RPA}.  Multiplying on the right by 
$\X^{-1}$, we have\footnote{For direct RPA, in which $\B$ is positive definite, 
$\X^{-1}$ exists, as proven in the appendix.  We must assume its existence for 
full RPA.}
\begin{equation}
\begin{pmatrix}
\hfill \A & \hfill \B \\  -\B & -\A
\end{pmatrix}
\begin{pmatrix}
\bm{1} \\ \T 
\end{pmatrix}
=
\begin{pmatrix}
\bm{1} \\ \T 
\end{pmatrix}
\R,
\label{RPATransform}
\end{equation}
where we have defined
\begin{subequations}
\begin{alignat}{1}
\T &= \Y \, \X^{-1},
\label{DefP}
\\
\R &= \X \, \w \, \X^{-1}.
\label{DefR}
\end{alignat}
\end{subequations}
As seen below, $\T = \Y \X^{-1}$ corresponds to the solution of Eqn. 
\ref{Riccati}.  Multiplying on the left by $( \T \quad  -\bm{1})$ yields 
\begin{equation}
\begin{pmatrix}
\T && -\bm{1}
\end{pmatrix}
\begin{pmatrix}
\hfill\A & \hfill\B \\ -\B & -\A
\end{pmatrix}
\begin{pmatrix}
\bm{1} \\ \T
\end{pmatrix}
= \begin{pmatrix}
\T && -\bm{1}
\end{pmatrix}
\begin{pmatrix}
\bm{1} \\ \T
\end{pmatrix}
\R.
\end{equation}
Carrying out the matrix multiplications, we see that this is just Eqn. 
\ref{Riccati}.  From Eqn. \ref{RPATransform}, we have
\begin{equation}
\A + \B \, \T = \R,
\label{ObtainR}
\end{equation}
whence
\begin{equation}
\mathrm{Tr}(\B \, \T) = \mathrm{Tr}(\R - \A) = \mathrm{Tr}(\w - \A).
\end{equation}
The direct ring-CCD correlation energy is thus equal to the direct RPA 
correlation energy.  The extra factor of 1/2 in the ring-CCD correlation 
energy on the right-hand-side of Eqn. \ref{Ec_rCCD} makes the correlation 
energy exact to lead order, and it has been argued that it should therefore be 
included in defining the full RPA correlation energy.  See Ref. 
\onlinecite{Oddershede} and references therein for discussion of this point.

In order to obtain $\X$ and $\Y$ once we have $\T$, we can use Eqn. 
\ref{ObtainR} to construct $\R$.  From Eqn. \ref{DefR}, we can diagonalize 
$\R$ to get $\X$.  Once we have $\X$ and $\T$, we simply use $\Y = \T \, \X$ 
to get $\Y$.

Direct RPA is commonly used in condensed matter physics, where the exchange 
terms are usually removed from the two-particle Hamiltonian (and treated as 
vertex corrections), and where typically semilocal DFT orbitals and orbital 
energies (\textit{i.e.} those coming from the local density approximation or a 
generalized gradient approximation) are used.  The exchange-correlation energy 
in such a scheme is given by
\begin{equation}
E_{xc} =  \tilde{E}_x^{HF} + E_c^{dRPA},
\end{equation}
where $\tilde{E}_x^{HF}$ is the Hartree-Fock-type exchange energy with the 
semilocal orbitals and where ``dRPA'' indicates direct RPA.  The pros and cons 
of keeping or neglecting vertex corrections in RPA correlation have been 
discussed in the literature.\cite{RubioRMP}

Given that both $\braket{ib}{aj}$ and $-t_{ij}^{ab}$ are positive definite for 
dRPA,\footnote{The latter is proven in the appendix.}  we can use Cholesky 
decomposition to write
\begin{subequations}
\begin{alignat}{1}
\braket{ib}{aj} &= \braket{ij}{ab} = u_{ia}^A \, u_{jb}^A,
\\
-t_{ij}^{ab}     &= \theta_{ia}^A \, \theta_{jb}^A,
\end{alignat}
\end{subequations}
where $A$ is to be summed.  This leads to the drCCD equation (Eqn. \ref{rCCD} 
with no exchange integrals) becoming
\begin{equation}
t_{ij}^{ab} = \frac{1}{\Delta \epsilon_{ij}^{ab}} \left(u_{ia}^A \, u_{jb}^A - u_{ia}^A \, u_{kc}^A \, \theta_{kc}^B  \, \theta_{jb}^B - \theta_{ia}^A \, \theta_{kc}^A \, u_{kc}^B \, u_{jb}^B + \theta_{ia}^A \, \theta_{kc}^A\,  u_{kc}^B \, u_{ld}^B \, \theta_{ld}^C \, \theta_{jb}^C\right)
\end{equation}
with
\begin{equation}
\Delta \epsilon_{ij}^{ab} = \epsilon_i + \epsilon_j - \epsilon_a - \epsilon_b.
\end{equation}
Defining
\begin{subequations}
\begin{alignat}{1}
M^{AB} &= \theta_{kc}^A \, u_{kc}^B,
\\
N^{AB} &= u_{kc}^A \, \theta_{kc}^B,
\end{alignat}
\end{subequations}
the construction of which scale as $\mathcal{O}(o v c^2)$ where 
$c = \mathrm{dim}\{A\}$, leads to
\begin{equation}
t_{ij}^{ab} = \frac{1}{\Delta \epsilon_{ij}^{ab}} \left(u_{ia}^A \, u_{jb}^A 
 - u_{ia}^A \, N^{AB} \, \theta_{jb}^B - \theta_{ia}^A \, M^{AB} \, \theta_{jb}^B + \theta_{ia}^A \, M^{AB} \, N^{BC} \, \theta_{jb}^C\right),
\end{equation}
which can be solved by fixed point iteration with DIIS\cite{DIIS} in 
$\mathcal{O}(o v c^2)$ operations.  Analytic energy gradients can also be 
carried out using the standard CC approach.\cite{ScuseriaSchaefer}

In the current framework, the cost of RPA is not much greater than that of 
MP2.  The atomic orbital to molecular orbital integral transformation needed 
to build $\braket{ib}{aj}$ scales as $\mathcal{O}(N^5)$ for $N$ atomic 
orbitals, and the Cholesky decomposition for dense $\braket{ib}{aj}$ and 
$t_{ij}^{ab}$ will scale worse than $\mathcal{O}(N^4)$.  However, transforming 
back into the atomic orbital basis (as in our AO-CC based 
formalism\cite{AOCC}) will yield algorithms that scale near-linearly for 
sparse enough matrices.\cite{CGDMS}

The connection between the symplectic eigenvalue problem (Eqn. \ref{RPA}) 
and its associated Riccati equation (Eqn. \ref{Riccati}) is textbook material 
in Optimal Control Theory (see, for example, Ref. \onlinecite{Zhou}).  
Sanderson\cite{Sanderson} seems to have been the first to document this 
connection in the context of RPA; however, he neither mentions coupled cluster 
theory nor the agreement of correlation energies between RPA and rCCD.  His 
assumption about commuting boson excitation operators leads to an RPA ground 
state representation that is correct only for two-electron 
systems.\cite{OstlundKarplus}

In summary, we have offered an analytic proof that the excitation amplitudes 
of an approximate CCD model are related to the eigenvectors of the RPA model 
by $\T = \Y \, \X^{-1}$, and that the ground state correlation energies of 
these two models are identical.  This connection also lets us establish an
$\mathcal{O}(N^4)$ algorithm for the RPA correlation energy in a CC framework 
thanks to the mathematical properties of the solution ($\T < \bm{0}$).

This work was supported by the National Science Foundation (CHE-0807194 and 
CCF-0634902) and the Welch Foundation (C-0036).  We thank Filipp Furche for 
providing benchmark numerical results of direct RPA correlation energies, and 
Georg Kresse for recently reviving our interest in this problem.

\appendix
\section{Mathematical Details}
We here prove several statements about the solution of Eqn. \ref{Riccati}.  

\subsection{Symmetry of $\Y^\mathrm{T} \, \X$}
We begin by showing that $\X^{\mathrm{T}} \, \Y = \Y^\mathrm{T} \, \X$.  Start 
with the RPA equation, Eqn. \ref{RPA}, and multiply on the left by 
$(\Y^{\mathrm{T}} \quad -\X^{\mathrm{T}})$ to get
\begin{equation}
\Y^\mathrm{T} \, \A \, \X + \X^\mathrm{T} \, \A \, \Y + \Y^\mathrm{T} \, \B \, \Y + \X^\mathrm{T} \, \B \, \X = (\Y^\mathrm{T} \, \X - \X^\mathrm{T} \, \Y) \w.
\end{equation}
Since the left-hand-side is symmetric, we have
\begin{equation}
(\Y^\mathrm{T} \, \X - \X^\mathrm{T} \, \Y) \, \w = \w \, (\X^\mathrm{T} \, \Y - \Y^\mathrm{T} \, \X).
\end{equation}
Defining $\S = \Y^\mathrm{T} \, \X - \X^\mathrm{T} \, \Y$, we thus have
\begin{equation}
\S \, \w + \w \, \S = \bm{0}.
\end{equation}
In indicial form, this is
\begin{equation}
S_{ij} (\omega_i + \omega_j) = 0.
\end{equation}
Since we have taken $\omega_i$ positive, we must have $\S = \bm{0}$, and hence 
$\X^{\mathrm{T}} \, \Y = \Y^\mathrm{T} \, \X$.

\subsection{Existence of $\X^{-1}$}
For positive definite $\B$ (true in direct RPA, but not in the full RPA), the 
existence of $\X^{-1}$ can be proven.  Suppose that $\X \, \z = \bm{0}$ for 
some vector $\z \ne \bm{0}$.  Multiplying both sides of the RPA equations by 
$\z$ would then give us
\begin{subequations}
\begin{alignat}{1}
\B \, \Y \, \z &= \X \, \w \, \z,
\\
-\A \, \Y \, \z &= \Y \, \w \, \z.
\end{alignat}
\end{subequations}
Since $\X^{\mathrm{T}} \, \Y = \Y^\mathrm{T} \, \X$, we would have
\begin{equation}
\z^{\mathrm{T}} \, \Y^{\mathrm{T}} \, \B \, \Y \, \z = \z^{\mathrm{T}} \, \Y^{\mathrm{T}} \, \X \, \w \, \z =  \z^{\mathrm{T}} \, \X^{\mathrm{T}} \, \Y^{\mathrm{T}}  \w \, \z = \bm{0}.
\end{equation}
Since $\B$ is positive definite, this implies that $\Y \, \z = \bm{0}$.  But 
this would mean that 
$\bigl(\begin{smallmatrix} \X \\ \Y \end{smallmatrix}\bigr) \,\z = \bm{0}$, 
contradicting the assumption that 
$\bigl(\begin{smallmatrix} \X \\ \Y \end{smallmatrix}\bigr)$ is of full rank 
made in writing the eigenvalue problem.

\subsection{Symmetry of $\T$}
Since $\X$ is nonsingular, and $\Y^\mathrm{T} \, \X = \X^\mathrm{T} \, \Y$,
we have
\begin{equation}
(\X^\mathrm{T})^{-1} \left( \Y^\mathrm{T} \, \X - \X^\mathrm{T} \, \Y \right) \X^{-1} = \bm{0}.
\end{equation}
Expanding the foregoing shows that
\begin{equation}
\T^{\mathrm{T}} - \T = \bm{0}.
\end{equation}

\subsection{Negative Definiteness of $\T$}
Since $\T$ is real and symmetric, we can diagonalize it with a unitary 
transformation $\mathbf{U}$:  $\T \, \mathbf{U} = \mathbf{U} \, \bm{\lambda}$. 
Multiplying the drCCD equation on the left by a particular eigenvector 
$\mathbf{U}_k^\dagger$ and on the right by $\mathbf{U}_k$, we get
\begin{equation}
\mathbf{U}_k^\dagger \, \B \, \mathbf{U}_k \left(1 + \lambda_k^2\right)  + 2 \, \mathbf{U}_k^{\dagger} \, \A \, \mathbf{U}_k \, \lambda_k = 0.
\end{equation}
When $\B$ and $\A$ are positive definite, as they are for direct RPA, we see 
that we must have $\lambda_k < 0$ for all $k$, and $\T$ is therefore negative 
definite.

\bibliography{RPA}

\end{document}